\documentclass[english,a4paper]{article}%
\usepackage[T1]{fontenc}
\usepackage[latin9]{inputenc}
\usepackage{amsmath}
\usepackage{graphicx}
\usepackage{amsfonts}
\usepackage{babel}
\usepackage{amssymb}%
\providecommand{\U}[1]{\protect\rule{.1in}{.1in}}
\providecommand{\U}[1]{\protect\rule{.1in}{.1in}}
\providecommand{\U}[1]{\protect\rule{.1in}{.1in}}
\providecommand{\U}[1]{\protect\rule{.1in}{.1in}}
\makeatletter
\providecommand{\U}[1]{\protect\rule{.1in}{.1in}}
\providecommand{\U}[1]{\protect\rule{.1in}{.1in}}
\providecommand{\U}[1]{\protect\rule{.1in}{.1in}}
\providecommand{\U}[1]{\protect\rule{.1in}{.1in}}
\providecommand{\U}[1]{\protect\rule{.1in}{.1in}}
\providecommand{\U}[1]{\protect\rule{.1in}{.1in}}
\providecommand{\U}[1]{\protect\rule{.1in}{.1in}}
\providecommand{\U}[1]{\protect\rule{.1in}{.1in}}
\providecommand{\U}[1]{\protect\rule{.1in}{.1in}}
\providecommand{\U}[1]{\protect\rule{.1in}{.1in}}
\providecommand{\U}[1]{\protect\rule{.1in}{.1in}}
\providecommand{\U}[1]{\protect\rule{.1in}{.1in}}

\makeatother
\begin{document}

\title{Quantum properties of a single beam splitter}
\author{F. Lalo\"{e}$^{a}$ and W.J.Mullin$^{b}$}
\maketitle

\begin{abstract}
When a single beam-splitter receives two beams of bosons described by Fock
states (Bose-Einstein condensates at very low temperatures), interesting
generalizations of the two-photon Hong-Ou-Mandel effect take place for larger
number of particles. The distributions of particles at two detectors behind
the beam splitter can be understood as resulting from the combination of two
effects, the spontaneous phase appearing during quantum measurement, and the
quantum angle. The latter introduces quantum ``population oscillations'',
which can be seen as a generalized Hong-Ou-Mandel effect, although they do not
always correspond to even-odd oscillations.

\end{abstract}
\date{}

{$^{a}$Laboratoire Kastler Brossel, ENS, UPMC, CNRS; 24 rue Lhomond, 75005
Paris, France }

$^{b}$Department of Physics, University of Massachusetts, Amherst,
Massachusetts 01003 USA



\begin{center}
********
\end{center}

\bigskip

Beam splitters are an essential component of many experiments designed to
observe quantum effects.\ They are involved in experimental and theoretical
schemes that both Helmut Rauch and Daniel Greenberger have studied.\ Indeed,
the famous neutron experiments of H.\ Rauch and colleagues
\cite{Rauch-neutron-interferometry, Rauch-livre} were made possible by the
realization of an appropriate device allowing neutron beams to be split into
two coherent beams, which can then be recombined and give rise to various
interesting quantum interference effects.\ The observation of equally famous
quantum GHZ (Greenberger, Horne and Zeilinger) violations of local realism
\cite{GHZ, GHZ-bis} may also require the use of photon beam splitters
\cite{GHZ-expts-1}.\ Still another example is given by the entanglement
swapping effect, which requires indistinguishable photons to be measured at
the output ports of a beam splitter \cite{Entanglement-swapping}. This list
is, of course, non-exhaustive.

Here, we come back to the basic properties of a single beam splitter and show
that, as simple as it may look, it already exhibits strong quantum properties.
Previous studies of the quantum properties of beam splitters include
Refs.\ \cite{Scully, Huttner, Luis}.\ Generally, beam splitters are used in
conditions where they receive particles one by one.\ Here we generalize the
discussion and consider the case where a beam splitter receives groups of
particles in its two input beams, described by Fock states of bosons.\ We will
study the effects of the \textquotedblleft quantum angle\textquotedblright,
which was introduced in the context of more elaborate interferometry
experiments involving several beam splitters, and leads to violations of local
realist BCHSH\ and GHZ inequalities \cite{Mullin-Laloe, Laloe-Mullin}. Holland
and Burnett have studied the quantum limits on the detection of small phase
shifts with interferometers involving two beam splitters
\cite{Holland-Burnett}; for this purpose they also study the distribution of
the relative phase of the two output beams at a single beam splitter, assuming
that the two incoming Fock states have equal populations (twin states).\ Here
we release this assumption and study the distributions of the number of
particles at the outputs.

The production of Fock states with photons is not an easy task, if not
impossible, except for a small number of photons; see for instance Ref.
\cite{Cosme} for a description of an experiment with two states containing two
photons. Coherent states are, of course, much easier to produce, even with a
small average number of photons (very small intensities), but they remain
fundamentally very different from Fock states.\ Fortunately, the phenomenon of
Bose-Einstein condensation in ultra-cold gases provides us with a method to
produce condensates thermally and, when repulsive interactions between the
atoms stabilize the condensate, there are good reasons to believe that its
state is well described by a Fock state.\ Thermal excitations are of course
always present, but they can be reduced very efficiently by reducing the
temperature.\ Moreover, the technique of Bragg scattering of atoms from
standing laser waves \cite{Pritchard} can be used to obtain efficient atomic
beam splitters \cite{Chu}, and even observe interferences with Bose-Einstein
condensates in interferometers with macroscopic arm separation \cite{Sackett}.

\section{Classical and quantum calculation}

\begin{figure}[h]
\centering \includegraphics[width=2.5in]{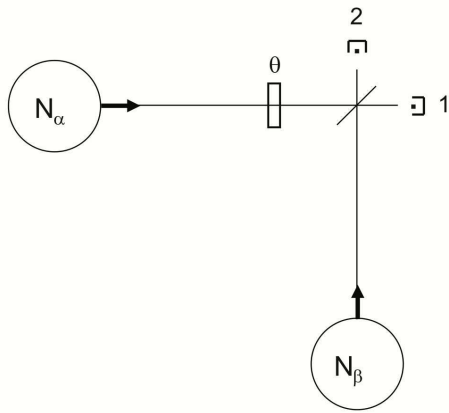}\caption{$N_{\alpha
},N_{\beta}$ bosons proceed from the sources to a beam splitter, followed by
two detectors 1 and 2, where $m_{1}$ and $m_{2}$ particles are detected. A
phase shift $\theta$ is inserted in the $\alpha$-arm for generality, but turns
out to play no role in the results. }%
\label{fig1-1}%
\end{figure}

The situation we consider is shown schematically in Fig.\ \ref{fig1-1}.\ We
will perform a quantum calculation but, as a point of comparison, we start
with a simple classical calculation.

\subsection{Classical model}

In classical optics, if two beams with equal intensities $I_{0}$ are sent to a
beam splitter with a relative phase $\lambda$, the output intensities $I_{1}$
and $I_{2}$ at the output detectors 1 and 2 are proportional to $\left[
1+\cos\left(  \lambda-\pi/2\right)  \right]  $ and $\left[  1-\cos\left(
\lambda-\pi/2\right)  \right]  $ (the $\pi/2$ arises because a phase shift
occurs at a reflection, but not at a transmission). If the phase $\lambda$ is
completely unknown, these expressions have to be summed over $\lambda$ between
$-\pi$ and $+\pi$; a well-known classical calculation then shows that the
distribution $P(I)$ of the random variables $I_{1,2}$ is given by:%
\begin{equation}
P_{class.}(I)\sim\frac{1}{\sqrt{I\left(  2I_{0}-I\right)  }}%
\label{distribution-classique}%
\end{equation}
If the input intensities are different, $I_{\alpha}$ and $I_{\beta}%
=x^{2}I_{\alpha}$, this calculation can easily be generalized. The two
intensities are now proportional to $\left[  1+r\cos\left(  \lambda
-\pi/2\right)  \right]  $ and $\left[  1-r\cos\left(  \lambda-\pi/2\right)
\right]  $, where:
\begin{equation}
r=\frac{2x}{1+x^{2}}=\frac{2\sqrt{I_{\alpha}I_{\beta}}}{I_{\alpha}+I_{\beta}%
}\leq1\label{def-r}%
\end{equation}
and (\ref{distribution-classique}) becomes:%
\begin{equation}
P_{class.}(I)=\frac{1}{\pi\sqrt{\left[  I-I_{\alpha}\left(  1-x\right)
^{2}/2\right]  \left[  -I+I_{\alpha}\left(  1+x\right)  ^{2}/2\right]  }%
}\label{inequal-intensities}%
\end{equation}
These expressions result from purely classical wave theory.

\begin{figure}[h]
\includegraphics[width=2.9in]{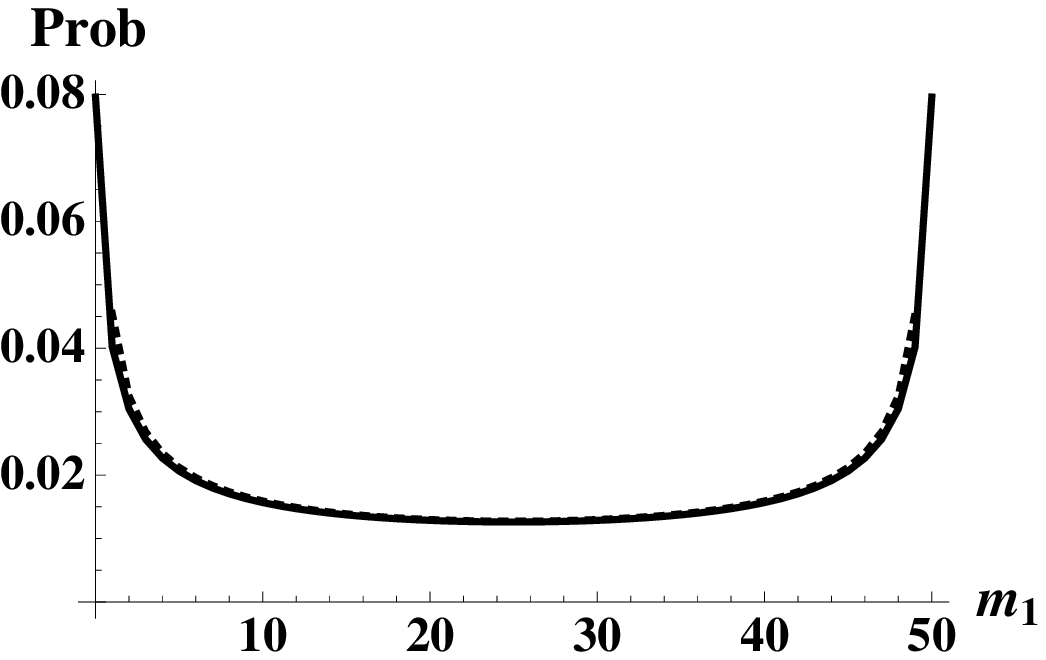}\quad{}%
\includegraphics[width=3.2in]{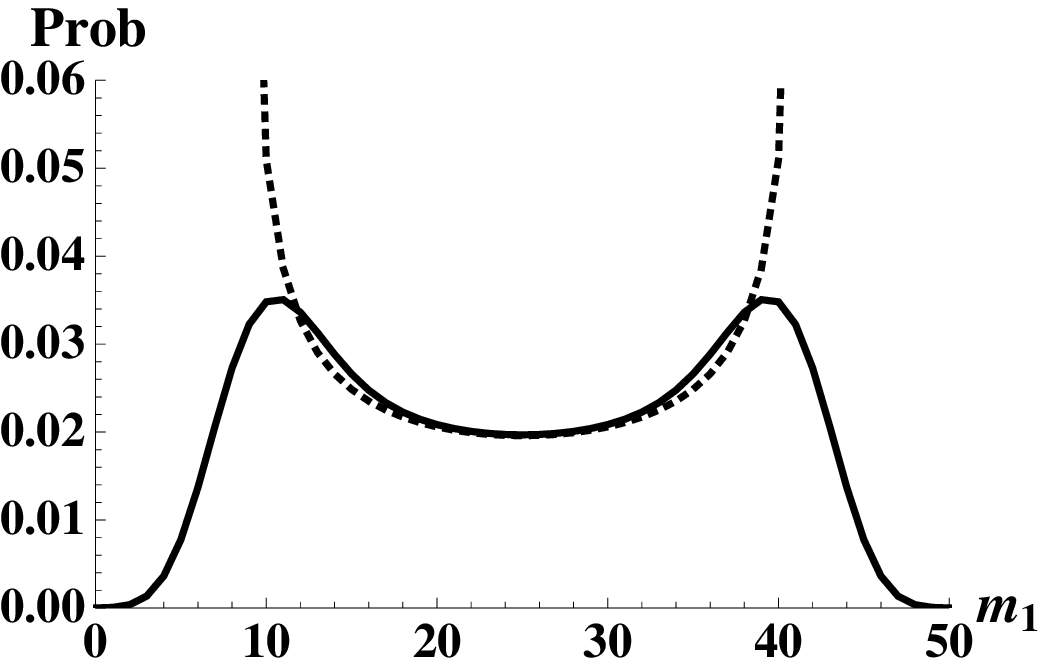}\caption{The left part (a) shows the
classical (dotted line) and semiclassical (full line) distributions as a
function of $m_{1}$ when the intensities of the input beams are equal
($x=r=1$) and when $m_{1} + m_{2} = 50$. Values of $m_{1}$ near the maximum
and the minimum are more likely to occur with this distribution. The right
part (b) shows the same distributions for the same total number of particles,
but when the intensities of the input beams are different (their ratio is
$6/44$).}%
\label{figNoLam}%
\end{figure}

Semi-classical expressions can be obtained by considering a flux of classical
particles reaching independently the beam splitter, each having a probability
$\left[  1+r\cos\left(  \lambda-\pi/2\right)  \right]  /2$ \ to go to detector
1, and a probability $\left[  1-r\cos\left(  \lambda-\pi/2\right)  \right]
/2$ to go to detector 2.\ The probability that, among a total of
$N\,$\ particles, $m_{1}$ will go to detector 1 and $m_{2}$ to detector 2
(with $m_{1}+m_{2}=N$) is then given by:%
\begin{equation}
P_{semi-class.}(m_{1},m_{2})=\frac{N!}{2^{N}~m_{1}!m_{2}!}\left[
1+r\cos\left(  \lambda-\pi/2\right)  \right]  ^{m_{1}}\left[  1-r\cos\left(
\lambda-\pi/2\right)  \right]  ^{m_{2}} \label{Pclass}%
\end{equation}
For Fock states, we expect that the relative phase $\lambda$ should be
completely random, so that this expression becomes:%
\begin{equation}
P_{semi-class.}^{Fock}(m_{1},m_{2})=\frac{N!}{2^{N}~m_{1}!m_{2}!}\int_{-\pi
}^{+\pi}\frac{d\lambda}{2\pi}~\left[  1+r\cos\lambda\right]  ^{m_{1}}\left[
1-r\cos\lambda\right]  ^{m_{2}} \label{Pclass-bis}%
\end{equation}
Figure \ref{figNoLam}-a shows an example of such a distribution for equal
intensities of the incoming beams, which reproduces the shape of the classical
distribution (\ref{distribution-classique}), with a minimum at the center and
maxima at the edges. Figure \ref{figNoLam}-b shows another example, assuming
now that the intensities of the two incoming beams are different, and that
their ratio is $6/44$. Because the interference effect between the two beams
can no longer be completely destructive, the distribution tends to concentrate
more towards the center or the curve (medium values of $m_{1}$).

An equivalent summation formulation more suitable for computations is found by
expanding the binomials $\left(  1\pm r\cos\lambda\right)  ^{m_{i}}$ and
integrating term by term.\ The result is:
\begin{equation}
P_{semi-class.}^{Fock}=K\left[  \sum_{p=0}^{m_{1}}\sum_{q=0}^{m_{2}}%
\frac{\left(  -1\right)  ^{q}~r^{p+q}~\left(  p+q\right)  !\left[  1+\left(
-1\right)  ^{p+q}\right]  }{2^{p+q}~\ p!\left(  m_{1}-p\right)  !~q!\left(
m_{2}-q\right)  !\left[  \left(  \frac{p+q}{2}\right)  !\right]  ^{2}}\right]
\label{approx}%
\end{equation}
where $K$ is a normalization factor.\ This formula was used, for instance, to
produce Fig. \ref{figNoLam} and the dotted line of Fig. \ref{figN2822}-b.

\subsection{Quantum calculation}

We give two calculations, for input beams described either by Fock states or
by coherent states.

\subsubsection{Fock states}

Before the beams of bosons cross the beam splitter, they are described by the
quantum state:%
\begin{equation}
\left\vert N_{\alpha},N_{\beta}\right\rangle =\frac{1}{\sqrt{N_{\alpha
}!N_{\beta}!}}a_{\alpha}^{\dagger N_{\alpha}}a_{\beta}^{\dagger N_{\beta}%
}\left\vert \text{0}\right\rangle \label{initialstate}%
\end{equation}
Our calculation that is essentially the same as that of \cite{Mullin-Laloe}
and \cite{Laloe-Mullin}. The destruction operators associated with the two
output beams (and detectors) are:%
\begin{equation}
a_{1}=\frac{1}{\sqrt{2}}\left(  e^{i\theta}a_{\alpha}+ia_{\beta}\right)
;\qquad a_{2}=\frac{1}{\sqrt{2}}\left(  ie^{i\theta}a_{\alpha}+a_{\beta
}\right)  \label{def-a}%
\end{equation}
The amplitude for finding $m_{1},m_{2}$ particles in the detectors given
sources with $N_{\alpha},N_{\beta}$ particles in the sources is:%
\begin{align}
C_{m_{1}m_{2}}(N_{\alpha},N_{\beta})  &  =\frac{1}{\sqrt{m_{1}!m_{2}%
!N_{\alpha}!N_{\beta}!}}\left\langle 0\right\vert a_{1}^{m_{1}}a_{2}^{m_{2}%
}a_{\alpha}^{\dagger N_{\alpha}}a_{\beta}^{\dagger N_{\beta}}\left\vert
0\right\rangle \nonumber\\
&  =\frac{\sqrt{N_{\alpha}!N_{\beta}!}}{\sqrt{m_{1}!m_{2}!}}\frac{e^{i\chi}%
}{\left(  \sqrt{2}\right)  ^{m_{1}+m_{2}}}\sum_{p,q}i^{q-p}\frac{m_{1}%
!}{p!(m_{1}-p)!}\frac{m_{2}!}{q!(m_{2}-q)!}\nonumber\\
&  \times\delta_{p+q,N_{\alpha}}\delta_{m_{1}+m_{2}-p-q,N_{\beta}}
\label{eq:C}%
\end{align}
where $\chi$ is a phase factor without physical relevance.\ Two methods of
calculation are now possible.

It is possible to replace the second $\delta$-function in Eq. (\ref{eq:C})
by:
\begin{equation}
\delta_{m_{1}+m_{2}-p-q,N_{\beta}}=\int\frac{d\phi}{2\pi}e^{i\phi(m_{1}%
+m_{2}-p-q-N_{\beta})} \label{delta}%
\end{equation}
to \ obtain:%

\begin{align}
C_{m_{1}m_{2}}(N_{\alpha}N_{\beta})  &  =\frac{e^{i\chi}}{2^{N}}\sqrt
{\frac{N_{\alpha}!N_{\beta}!}{m_{1}!m_{2}!}}\int_{-\pi}^{\pi}\frac{d\phi}%
{2\pi}e^{-iN_{\beta}\phi}\nonumber\\
&  \times\left(  e^{i\theta}+ie^{i\phi}\right)  ^{m_{1}}\left(  ie^{i\theta
}+e^{i\phi}\right)  ^{m_{2}} \label{eq:Rphi}%
\end{align}
The square the modulus of this expression contains an integral over two
variables $\varphi$ and $\varphi^{\prime}$;\ if we make the changes of
variables:%
\begin{equation}
\lambda=\frac{\phi+\phi^{\prime}+\pi}{2}-\theta\text{ \ \ \ ;\ \ \ \ \ }%
\Lambda=\frac{\phi-\phi^{\prime}}{2} \label{def-lambda}%
\end{equation}
we find for the probability the expression:%
\begin{align}
P(m_{1},m_{2})  &  =\frac{N_{\alpha}!N_{\beta}!}{m_{1}!m_{2}!}\int_{-\pi}%
^{\pi}\frac{d\lambda}{2\pi}\int_{-\pi}^{\pi}\frac{d\Lambda}{2\pi}\cos\left[
\left(  N_{\alpha}-N_{\beta}\right)  \Lambda\right] \nonumber\\
&  \left[  \cos\Lambda+\cos\lambda\right]  ^{m_{1}}\left[  \cos\Lambda
-\cos\lambda\right]  ^{m_{2}} \label{eq:QAform}%
\end{align}
(note that the phase shift $\theta$ has disappeared from this result). Assume
for a moment that the $\Lambda$ can be replaced by $0$ in the three cosines
that contain it.\ Then $\Lambda$ disappears, and we are left with an
expression that is identical to (\ref{Pclass-bis}) with $r=1$, except for
normalization factors.\ We therefore see that $\lambda\,$ (or, more precisely,
$\lambda+\pi/2$) plays the role of the classical relative phase of the two
sources; this phase is averaged over $2\pi$, which is normal since the phase
in a Fock state is completely undetermined. For this reason, we will call
$\lambda$ the classical phase angle, and $\Lambda$ the quantum angle; we study
in more detail below how the non-zero values of $\Lambda$ introduce quantum effects.

Another method is to use the $\delta-$functions to eliminate the summation
variable $q$ in Eq. (\ref{eq:C}) and then square the result.\ We then find:%
\begin{equation}
P(m_{1},m_{2})=\frac{m_{1}!m_{2}!N_{\alpha}!N_{\beta}!}{2^{N}}\left[
\sum_{p=0}^{m_{1}}\frac{(-1)^{p}}{p!(m_{1}-p)!(N_{\alpha}-p)!(p+m_{2}%
-N_{\alpha})!}\right]  ^{2} \label{autre-expression}%
\end{equation}
This expression is more convenient than (\ref{eq:QAform}) for accurate
numerical calculations.

\subsubsection{Coherent states}

We replace the ket (\ref{initialstate}) by a product of coherent input states:%
\begin{equation}
\left\vert \Psi\right\rangle \sim\sum_{n_{\alpha}}\sum_{n_{\beta}}\frac
{1}{n_{\alpha}!}\frac{1}{n_{\beta}!}~\left[  E_{\alpha}~a_{\alpha}^{\dagger
}\right]  ^{n_{\alpha}}\left[  E_{\beta}~a_{\beta}^{\dagger}\right]
^{n_{\beta}}\left\vert \text{0}\right\rangle =\exp\left[  E_{\alpha}%
~a_{\alpha}^{\dagger}+E_{\beta}~a_{\beta}^{\dagger}\right]  \left\vert
\text{0}\right\rangle \label{psi-coherent}%
\end{equation}
where $E_{\alpha}$ and $E_{\beta}$ are complex number defining the intensity
$I_{\alpha,\beta}=\left\vert E_{\alpha,\beta}\right\vert ^{2}$ and the phases
$\varphi_{\alpha,\beta}$ of the incoming fields. In this expression, we can
replace the creation operators by their expressions obtained from
(\ref{def-a}) and obtain:%
\begin{equation}
\exp\left[  E_{\alpha}~a_{\alpha}^{\dagger}+E_{\beta}~a_{\beta}^{\dagger
}\right]  \left\vert \text{0}\right\rangle =\exp\left[  \frac{E_{\alpha
}e^{i\theta}+iE_{\beta}}{\sqrt{2}}~a_{1}^{\dagger}+\frac{iE_{\alpha}%
e^{i\theta}+E_{\beta}}{\sqrt{2}}~a_{2}^{\dagger}\right]  \left\vert
\text{0}\right\rangle \label{exp}%
\end{equation}
Therefore, the output state is a product of coherent states as well, with
amplitudes of the fields given by $\left(  E_{\alpha}+iE_{\beta}\right)
/\sqrt{2}$ and $\left(  iE_{\alpha}+E_{\beta}\right)  /\sqrt{2}$, which
correspond exactly to the classical formulas. The operators contained in the
exponential commute.\ By expanding it into a series as in (\ref{psi-coherent}%
), we obtain the probability to measure $m_{1}$ bosons at output 1 and $m_{2}$
at output 2 as a product of Poissonian distributions:%
\begin{equation}%
\begin{array}
[c]{cc}%
P(m_{1},m_{2}) & \displaystyle\sim\frac{1}{m_{1}!}\frac{1}{m_{2}!}\left\vert
\frac{E_{\alpha}e^{i\theta}+iE_{\beta}}{\sqrt{2}}\right\vert ^{2m_{1}%
}\left\vert \frac{iE_{\alpha}e^{i\theta}+E_{\beta}}{\sqrt{2}}\right\vert
^{2m_{2}}\\
& \displaystyle\sim\frac{\left(  I_{\alpha}+I_{\beta}\right)  ^{m_{1}+m_{2}}%
}{2^{m_{1}+m_{2}}~m_{1}!m_{2}!}\left[  1+r\cos\left(  \lambda-\pi/2\right)
\right]  ^{m_{1}}\left[  1-r\cos\left(  \lambda-\pi/2\right)  \right]
^{m_{2}}%
\end{array}
\label{Poisson}%
\end{equation}
where $\lambda=\varphi_{\alpha}- \varphi_{\beta}+\theta$. The result is
therefore very similar to Eq.\ (\ref{Pclass}), as well as to (\ref{Pclass-bis}%
) if we assume that the phases of the incoming coherent beams are random.

\subsection{The generalized beam splitter theorem}

The properties of the distributions obtained in (\ref{eq:QAform}) and
(\ref{eq:QAform}) yield the generalized Hong-Ou-Mandel theorem as we see next.

\subsubsection{Calculation}

The second line in Eq. (\ref{eq:Rphi}) can be factored to produce:%
\begin{equation}
\left(  e^{i\theta}+ie^{i\phi}\right)  ^{m_{1}}\left(  ie^{i\theta}+e^{i\phi
}\right)  ^{m_{2}}=(i)^{m_{1}}\;2^{N}\;e^{i\bar{\phi}N/2}\;Q(\bar{\phi})
\end{equation}
where:%
\begin{equation}
Q(\bar{\phi})=\left(  \cos\frac{\bar{\phi}}{2}\right)  ^{m_{1}}\left(
\sin\frac{\bar{\phi}}{2}\right)  ^{m_{2}}%
\end{equation}
with:%
\begin{equation}
\bar{\phi}\equiv\phi-\theta+\frac{\pi}{2}%
\end{equation}
An exact result then is:%

\begin{align}
C_{m_{1}m_{2}}(N_{\alpha}N_{\beta})  &  = 2^{N/2}I\sqrt{\frac{N_{\alpha
}!N_{\beta}!}{m_{1}!m_{2}!}}\int_{-\pi}^{\pi}\frac{d\bar{\phi}}{2\pi
}e^{i(N_{\alpha}-N_{\beta})\bar{\phi}/2}\nonumber\\
&  \times\left(  \cos\frac{\bar{\phi}}{2}\right)  ^{m_{1}}\left(  \sin
\frac{\bar{\phi}}{2}\right)  ^{m_{2}} \label{Exactcossin}%
\end{align}

If $N_{\alpha}=N_{\beta}=1$, we are in the situation of the Hong-Ou-Mandel
effect \cite{Hong-Ou-Mandel}.\ When two photons fall on a beam splitter from
two symmetrical directions, it is known that a quantum interference effect
prevents them from leaving the beam splitter separately; they always leave
together (in the same direction).\ Here we obtain a direct generalization of
this theorem: if an equal number of particles approaches from each side to
meet at the beam splitter, an even number must emerge from each side.\ This is
because, if $N_{\alpha}=N_{\beta}$ in the probability amplitude
(\ref{Exactcossin}), we have $N=m_{1}+m_{2}$ even, in which case $m_{1}$ and
$m_{2}$ are both even or both odd; but, if $m_{2}$ is odd, the integral is
over an odd function and therefore vanishes.

\subsubsection{A Gaussian fit}

\begin{figure}[h]
\centering \includegraphics[width=3in]{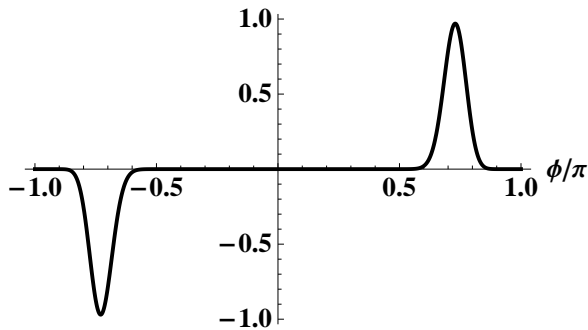}\caption{A plot of
$Q(\phi)$ for $m_{1}=17$ and $m_{2}=83$ and the phase $\theta=\pi/2$. The
peaks are at $\phi_{0}=\pm0.73 \; \pi$ (the phase choice gives symmetrical
peaks about zero). The relative sign of the two peaks is $(-1)^{m_{2}}$. The
peaks are normalized to unit height here. }%
\label{figRofphi}%
\end{figure}

A plot of the second line in Eq. (\ref{Exactcossin}) is shown in Fig.
\ref{figRofphi}. We see from the figure that a (double) Gaussian fit to
$Q(\bar{\phi})$ is likely to be a good approximation. We set the first
derivative of the logarithm of $Q(\bar{\phi})$ to zero, which gives the maxima
positions at $\pm\phi_{0}$ where:
\begin{equation}
\phi_{0}=2\arccos\left(  \sqrt{\frac{m_{1}}{N}}\right)  \label{eq:phi0}%
\end{equation}
and also find that the second derivative there is $-2N.$ The peak at
$-\phi_{0}$ is negative if $m_{2}$ is odd. Hence we get the approximation:%
\begin{equation}
Q(\bar{\phi})=\left(  \frac{m_{1}}{N}\right)  ^{m_{1}/2}\left(  \frac{m_{2}%
}{N}\right)  ^{m_{2}/2}\left[  e^{-\frac{N}{4}(\bar{\phi}-\phi_{0})^{2}%
}+(-1)^{m_{2}}e^{-\frac{N}{4}(\bar{\phi}+\phi)^{2}}\right]
\end{equation}
where the prefactor to the Gaussians comes from $Q(\phi_{0})$ upon use of
$\cos^{2}\phi_{0}/2=m_{1}/M$ and $\sin^{2}\phi_{0}/2=m_{2}/M$. In Eq.
(\ref{Exactcossin}) we then Fourier transform the Gaussians; doing these
integrals and squaring gives the probability:
\begin{align}
P_{m_{1}m_{2}}(N_{\alpha}N_{\beta})  &  =\frac{2^{N+2}N_{\alpha}!N_{\beta}%
!}{\pi N^{N+1}}\frac{m_{1}^{m_{1}}m_{2}^{m_{2}}}{m_{1}!m_{2}!}e^{-\frac{1}%
{2N}\left(  N_{\alpha}-N_{\beta}\right)  ^{2}}\nonumber\\
&  \times%
\begin{cases}
\cos\left[  \left(  N_{\alpha}-N_{\beta}\right)  \frac{\phi_{0}}{2}\right]
^{2} & \text{~~~~~~~for $m_{2}$~~even}\\
\sin\left[  \left(  N_{\alpha}-N_{\beta}\right)  \frac{\phi_{0}}{2}\right]
^{2} & \text{~~~~~~~for $m_{2}$~~odd}%
\end{cases}
\label{eq:almostthere}%
\end{align}

An interesting further approximation uses the Stirling formula for the
$m_{i}!$. We have%
\begin{equation}
m_{i}!\simeq\sqrt{2\pi}m_{i}^{m_{i}+\frac{1}{2}}e^{-m_{i}}%
\end{equation}
The factor of $1/2$ in the exponent is usually dropped, but is actually
important in our case in giving a characteristic shape to the probability
curves. Moreover it makes the Stirling formula accurate to within a few
percent for $m_{i}>2.$ The result is then:%
\begin{align}
P_{m_{1}m_{2}}(N_{\alpha}N_{\beta})  &  =\frac{2^{N+1}N_{\alpha}!N_{\beta}%
!}{\pi^{2}N^{N+1}}\frac{e^{-N}}{\sqrt{m_{1}m_{2}}}e^{-\frac{1}{2N}\left(
N_{\alpha}-N_{\beta}\right)  ^{2}}\nonumber\\
&  \times%
\begin{cases}
\cos\left[  \left(  N_{\alpha}-N_{\beta}\right)  \frac{\phi_{0}}{2}\right]
^{2} & \text{~~~~~~~for $m_{2}$~~even}\\
\sin\left[  \left(  N_{\alpha}-N_{\beta}\right)  \frac{\phi_{0}}{2}\right]
^{2} & \text{~~~~~~~for $m_{2}$~~odd}%
\end{cases}
\label{eq:ApproxP}%
\end{align}
Note the characteristic $\sqrt{m_{1}m_{2}}$ form in the denominator stemming
here from the Stirling formula. This formula matches the exact result in Eq.
(\ref{autre-expression}) very accurately for $m_{i}>2.$

\section{Physical discussion}

\subsection{Few bosons}

We now study in more detail the consequences of the rule according to which
even numbers of particles emerge from the beam splitter if an equal number
impinge on each side.\ For this purpose, we use Eq. (\ref{autre-expression})
to calculate the distribution corresponding to the various possible numbers of
particles detected at 1 and 2.

\begin{figure}[h]
\centering \includegraphics[width=3in]{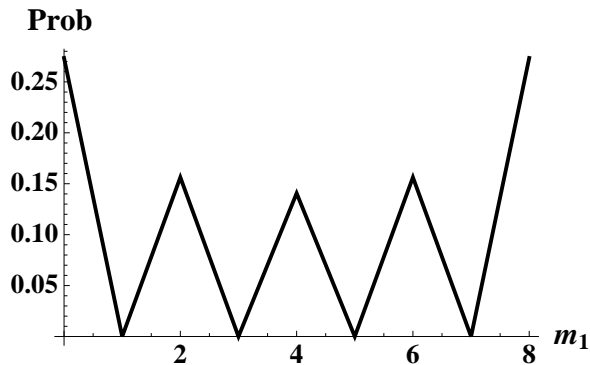}\caption{The probability for
$N_{\alpha}=4,\text{~}N_{\beta}=4$ illustrating the rule that, if an even
number of particles enters each side of the beam splitter, an even number must
emerge from each side.}%
\label{figNeq8}%
\end{figure}

Fig. \ref{figNeq8} shows four particles entering each side of the beam
splitter.\ As expected, only an even number can emerge on each side, which
explains the zeroes of the curve.\ Comparing with Fig. \ref{figNoLam}
immediately indicates that these cancellations superimpose strong quantum
oscillations onto the classical intensity distribution. We have a sort of
combination of a classical average over a phase $\lambda$ with rapid
variations created by the Hong-Ou-Mandel effect.

\begin{figure}[h]
\centering \includegraphics[width=3in]{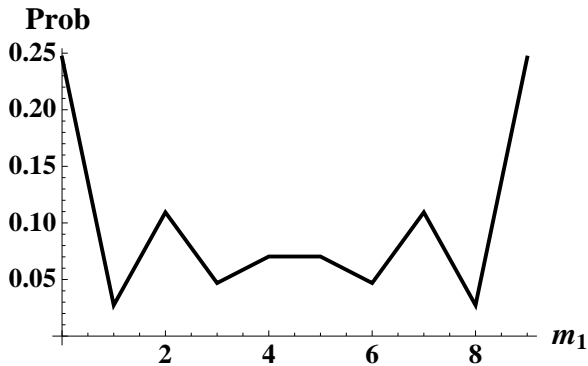}\caption{The probability for
$N_{\alpha}=4,\text{~}N_{\beta}=5$ . The probability no longer vanishes for
odd $m_{1}$ but oscillations remain. Now, of course, odd $m_{1}$ implies even
$m_{2}$ and vice versa.}%
\label{figNeq9}%
\end{figure}

For $N_{\alpha}=4,\text{~}N_{\beta}=5$, the rule no longer applies, but the
calculation of the distribution can still be done.\ The result is shown in
Fig. \ref{figNeq9}, which again contains odd-even variations and oscillations,
even if the probability does not vanish for any value of the $m$'s.\ In this
case also, we have the superposition of an average curve, which can be
understood in terms of a classical phase, and of an oscillation that can be
seen as a generalized Hong-Ou-Mandel effect.

\subsection{More particles}

\begin{figure}[h]
\centering \includegraphics[width=3in]{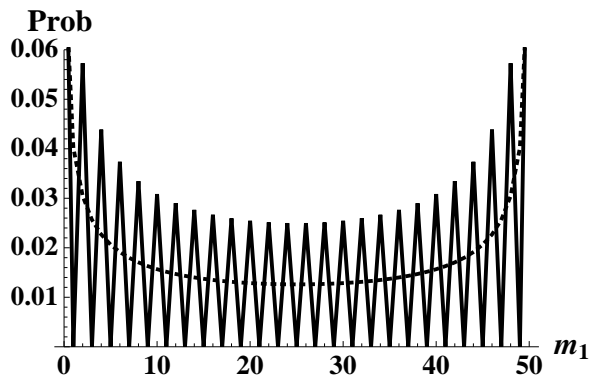}\caption{The probability for
$N_{\alpha}=\text{~}N_{\beta}=25$ . The probability vanishes for odd $m_{1}.$
The graph has been cut off at $m_{1}=0,50$ where it is about twice as high.
The variation with $1/\sqrt{m_{1}m_{2}}$ is evident. The dotted line shows the
corresponding semi-classical distribution}%
\label{figNeq50}%
\end{figure}

For larger values of $N$, the even rule is shown in Fig. \ref{figNeq50}, where
the variation of the probability with $m_{1}$ is plotted for $N_{\alpha
}=N_{\beta}=25.$ The characteristic probability variation $1/\sqrt{m_{1}m_{2}%
}$ given in Eq. (\ref{eq:ApproxP}) is visible, but again with strong quantum
oscillations. In this case in Eq. (\ref{eq:ApproxP}) the factor $\sin\left[
\left(  N_{\alpha}-N_{\beta}\right)  \phi_{0}/2\right]  ^{2}$ vanishes to
satisfy the even rule.

\begin{figure}[h]
\centering \includegraphics[width=3in]{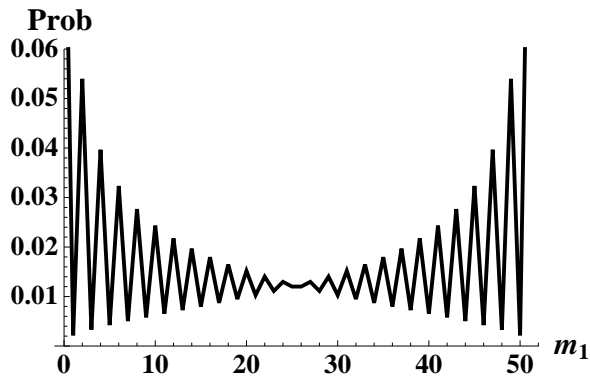}\caption{The probability for
$N_{\alpha}=26,\text{~}N_{\beta}=25$ . }%
\label{figNeq51}%
\end{figure}

Suppose now we have slightly different source populations, $N_{\alpha}=26$ and
$N_{\beta}=25$.\ The result is shown in Fig. \ref{figNeq51}. When $N_{\alpha
}-\text{~}N_{\beta}=1$ we can find explicitly that:
\begin{align}
\cos\left[  \left(  N_{\alpha}-N_{\beta}\right)  \frac{\phi_{0}}{2}\right]
^{2}  &  =\cos\left(  \frac{\phi_{0}}{2}\right)  ^{2}=\frac{m_{1}}%
{N}\nonumber\\
\sin\left[  \left(  N_{\alpha}-N_{\beta}\right)  \frac{\phi_{0}}{2}\right]
^{2}  &  =\sin\left[  \frac{\phi_{0}}{2}\right]  ^{2}=\frac{N-m_{1}}{N}%
\end{align}

\noindent The probability oscillates between these two values,
modulated by the $1/\sqrt{m_{1}m_{2}}$ factor, and changes over from maxima at
even to odd at $m_{1}=25$. It is interesting to see that, in this case, the
quantum oscillations vanish at the center of the distribution, but remain very
pronounced at both sides.

\begin{figure}[h]
\centering \includegraphics[width=3in]{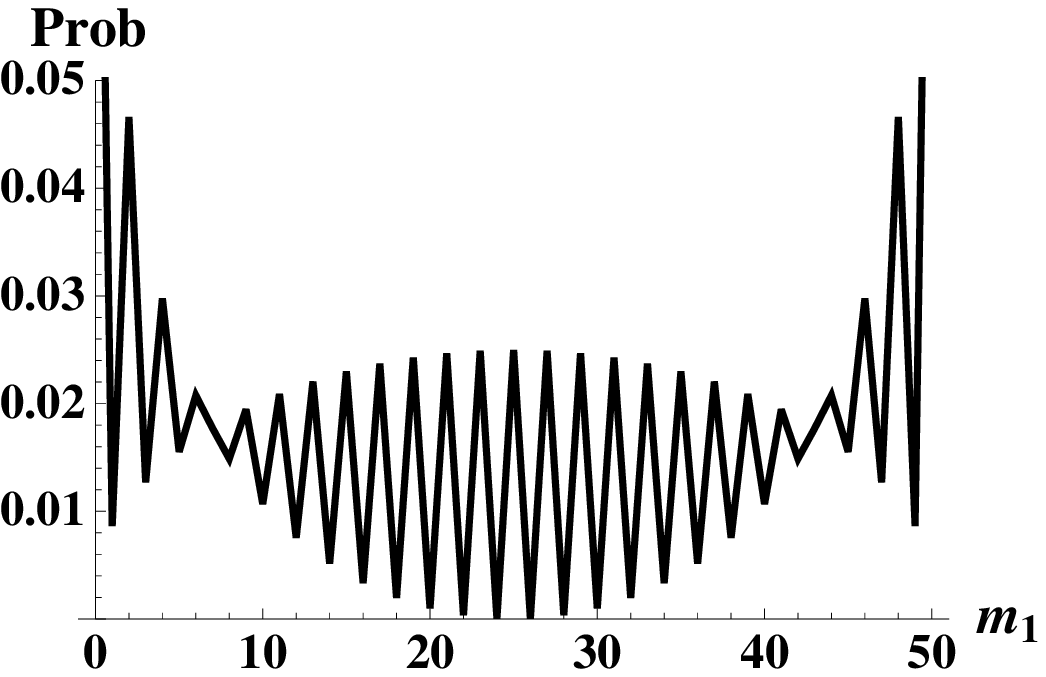}\caption{The probability for
$N_{\alpha}=26,\text{~}N_{\beta}=24$ . }%
\label{figN2624}%
\end{figure}

Larger population imbalances in the sources, at constant sum $N=50$, result in
even more complicated behavior.\ What happens for instance if $N_{\alpha
}-N_{\beta}=2$ is shown in Fig. \ref{figN2624}. We can then show that:
\begin{align}
\cos\left[  \left(  N_{\alpha}-N_{\beta}\right)  \frac{\phi_{0}}{2}\right]
^{2}  &  =\cos\left(  \phi_{0}\right)  ^{2}=\left(  \frac{2m_{1}}{N}-1\right)
^{2}\nonumber\\
\sin\left[  \left(  N_{\alpha}-N_{\beta}\right)  \frac{\phi_{0}}{2}\right]
^{2}  &  =\sin\left[  \phi_{0}\right]  ^{2}=4\frac{m_{1}(N-m_{1})}{N^{2}}%
\end{align}
The probability oscillates between these two curves with \textquotedblleft
nodes\textquotedblright\ at $m_{1}=7$ and $43$ corresponding to the crossing
of the two quadratic curves. The nodes are actually a consequence of the
discrete character of $m_{1}$: if $m_{1}$ is replaced by a continuous variable
in Eq. (\ref{Exactcossin}), then the probability distribution becomes an
oscillating function with a slowly varying amplitude. If the maxima and minima
occur near integer values of $m_{1}$, they remain very visible in the discrete
version of the distribution, resulting in antinodes; but, if they occur near
half integer values, the oscillations disappear in the discrete version,
resulting in nodes. This ``stroboscopic effect'' also explains the minimum of
oscillations at the center of Fig. \ref{figNeq51}.

\begin{figure}[h]
\includegraphics[width=3in]{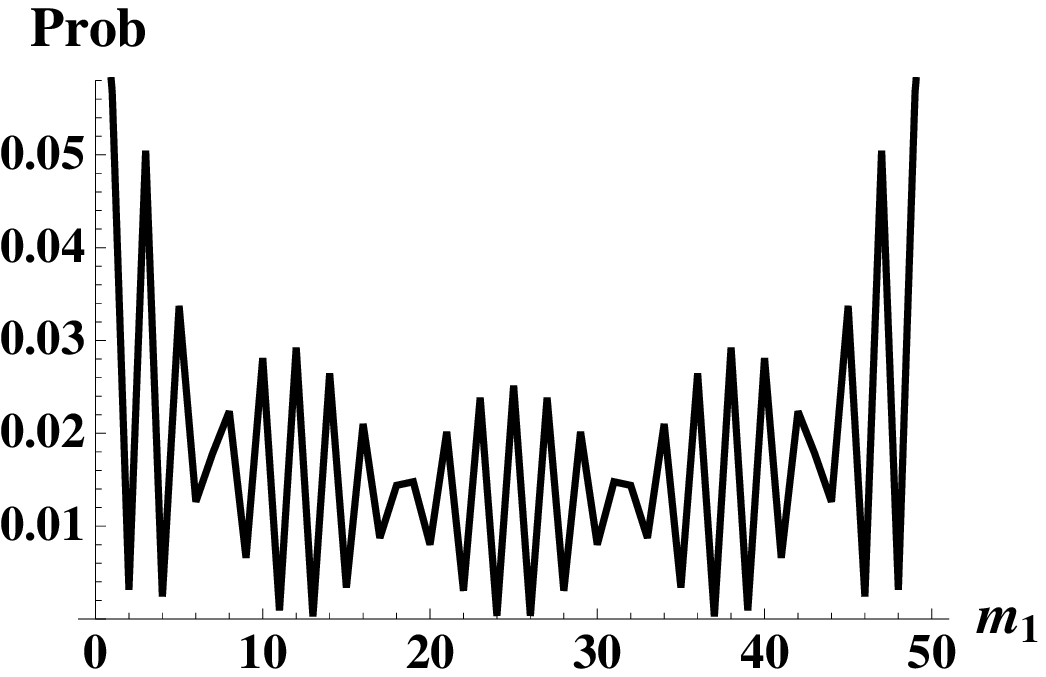}\quad{}%
\includegraphics[width=7cm]{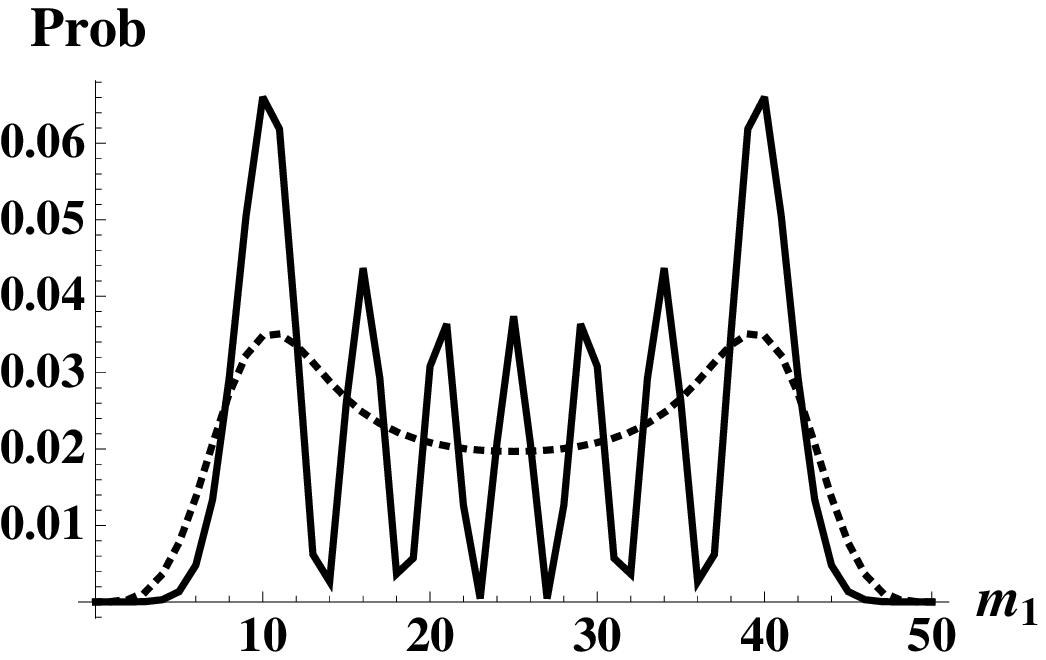}\caption{The left part (a) shows the
probability distribution for $N_{\alpha}=28,\text{~}N_{\beta}=22$. The right
part (b) shows the same curve for $N_{\alpha}=44,\text{~}N_{\beta}=6$ (full
line) and, as a point or comparison (dotted line) the classical curve of Fig.
\ref{figNoLam}-b.}%
\label{figN2822}%
\end{figure}

The ``beating wavelength'' becomes shorter as $N_{\alpha}-N_{\beta}$ becomes
larger. The case $N_{\alpha}=28,\text{~}N_{\beta}=22$ is shown in Fig.
\ref{figN2822}-a; the case $N_{\alpha}=44,\text{~}N_{\beta}=6$ is shown in
Fig. \ref{figN2822}-b. In this case, the fast oscillations of the generalized
Hong-Ou-Mandel type have disappeared and have become significantly slower,
which presumably makes them easier to observe experimentally.

\subsection{Role of the quantum angle}

If $N_{\alpha}=N_{\beta}$, we can compare the semi-classical expression, Eq.
(\ref{Pclass-bis}) with $r=1$, to the quantum result Eq.\ (\ref{eq:QAform}).
We see that, instead of classical probabilities $\left[  1\pm\cos
\lambda\right]  $, the quantum expression contains quasi-probabilities
$\left[  \cos\Lambda\pm\cos\lambda\right]  $, which can take negative values
when the quantum angle $\Lambda$ does not vanish. This introduces quantum
effects in a way that is reminiscent of quantum effects arising from negative
values of the Wigner transform. The quantum angle is responsible for the
population oscillations introduced by the beam splitter.

When $N_{\alpha}\neq N_{\beta}$, Eq.\ (\ref{eq:QAform}) shows that the quantum
angle also controls the effect of population imbalance.\ The quantum formula,
instead of including in the quasi-probabilities a factor:
\begin{equation}
r=\frac{2\sqrt{N_{\alpha}N_{\beta}}}{N_{\alpha}+N_{\beta}}%
\end{equation}
contains inside the integral an oscillating function $\cos\left[  \left(
N_{\alpha}-N_{\beta}\right)  \Lambda\right]  $. Figs. \ref{figN2624} and
\ref{figN2822} illustrate the effects of population imbalance.

To see what happens when the effect of $\Lambda$ is cancelled, let us set
$\Lambda=0$ in Eq. (\ref{eq:QAform}). We write $(1\pm\cos\lambda)$ in terms of
sine and cosine of the half angle, expand these in exponentials, expand the
binomials, and integrate. The result is:%
\begin{equation}
P_{\Lambda=0}(m_{1},m_{2})=\frac{N!(2m_{1})!(2m_{2})!}{4^{N}m_{1}!m_{2}%
!}\left[  \sum_{p=0}^{2m_{1}}\frac{(-1)^{m_{1}+p}}{p!(2m_{1}%
-p)!(N-p)!(p-2m_{1}+N)!}\right]  \label{P-Lambda=0}%
\end{equation}
No oscillation then takes place; for instance, the case of $N_{\alpha
}=N_{\beta}=25$ was already shown in Fig. \ref{figNoLam}.

\subsection{Pair-probability formulation}

When $N_{\alpha}=N_{\beta}=1$, the Hong-Ou-Mandel result is that the
probabilities of having $n$ particles in the detector 1 and and $2-n$ in
detector 2 is:
\begin{equation}
\mathcal{P}_{n}=\frac{1}{2}[1+(-1)^{n}]
\end{equation}
with $n=0,1,2$. Now, a natural question is: can we consider that the
distributions obtained above can be interpreted as those that one would be
obtained by repeating the Hong-Ou-Mandel experiment a sufficient number of
times, and accumulating the counts in each detector?

\begin{figure}[h]
\centering \includegraphics[width=3in]{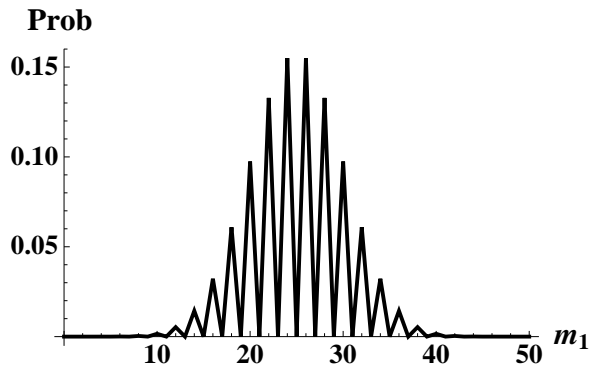}\caption{The pair probability
for $N_{\alpha}=25,\text{~}N_{\beta}=25$. }%
\label{figPairProb}%
\end{figure}

First consider the case when $N_{\alpha}=N_{\beta}=N/2.$ We pick the first
pair, and it produces either 2 particles on the left or none. Then the second
pair does the same. We continue until we have $m_{1}/2$ pairs on the left and
$m_{2}/2$ pairs on the right. We consider that, on the left we have $m_{1}/2$
filled pair slots and $m_{2}/2$ empty pair slots, which could have been
selected in any order. We interchange the slots, while not counting
interchanges of the empty slots among themselves and the filled among
themselves. We have then $(N/2)!/[(m_{1}/2)!(m_{2}/2)!]$ different ways for
getting the $(m_{1},m_{2})$ probability, which is:
\begin{equation}
P_{m_{1},m_{2}}^{(pair)}\sim\frac{(N/2)!}{(m_{1}/2)!(m_{2}/2)!}\left[
(1+(-1)^{m_{1}})\right]
\end{equation}
where the last factor ensures that there are an even number of particles on
each side. For 50 particles the result is shown in Fig. \ref{figPairProb}.

We see that the distribution obtained in this way still has the expected
odd-even behavior (actually almost by construction), but that it does not
reproduce the characteristic $1/\sqrt{m_{1}m_{2}}$ shape for the envelope.
Instead, it tends to concentrate the most likely results around $m_{1}\simeq
m_{2}\simeq N/2$, which is natural: if we have independent scattering events
of pairs into either channel, chosen randomly, one expects that the most
likely values will be equal in both outputs.

By contrast, the curve of Fig.\ \ref{figNeq50} has its maxima at $m_{1}=0$ and
$m_{1}=N$, which is a completely different behavior, and indicates the effect
of bosonic statistics (bunching into one channel).\ The distribution is no
longer peaked at the center but spreads towards both sides.\ As
Fig.\ \ref{figNoLam} shows, this behavior can be explained if we add a new
ingredient, a relative phase.\ But, since Fock sources do not have any initial
phase, this can be understood as a result of a spontaneous choice of a
relative phase by the two sources under the effect of quantum measurement
\cite{Laloe-Mullin}; since this phase is completely unknown, an average over
all possible values is taken.

\begin{figure}[h]
\centering \includegraphics[width=3in]{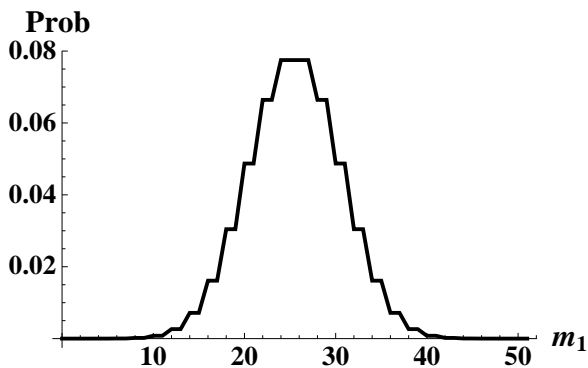}\caption{The pair
probability for $N_{\alpha}=26,\text{~}N_{\beta}=25$. }%
\label{figPairPlusSing}%
\end{figure}

Now, if we have an excess of particles on one side, for example, $N_{\alpha
}=N_{\beta}+\mathcal{N}$, then we can assume each $\beta$ particle is paired
with an $\alpha$ particle and the $\mathcal{N}$ extras appear anywhere in the
sequence of selections as singles. On the left side, we have $N_{\beta}$ pair
slots with $f$ of them filled and $e$ of them empty. Of the single slots on
the left, $s$ are filled and $o$ are open. Then we have
\begin{align}
N_{\beta}  &  =f+e\nonumber\\
N_{\alpha}-N_{\beta}  &  =s+o\nonumber\\
m_{1}  &  =2f+s
\end{align}
These can be solved to give:%
\begin{align}
f  &  =\frac{1}{2}(m_{1}-s)\nonumber\\
o  &  =N_{\alpha}-N_{\beta}-s\nonumber\\
e  &  =N_{\beta}-\frac{1}{2}(m_{1}-s)
\end{align}
Then among the pairs and singles on the left we can rearrange in
$f+e+s+o=N_{\alpha}$ total ways with rearrangements among the same kind of
slots not counting to give a probability
\begin{equation}
P_{m_{1}m_{2}}\sim\sum_{s=0}^{N_{\alpha}-N_{\beta}}\frac{(N_{\alpha}%
)!}{\left(  \frac{m_{1}-s}{2}\right)  !(N_{\alpha}-N_{\beta}-s)!s!\left(
N_{\beta}-\frac{1}{2}(m_{1}-s)\right)  !}\left[  (1+(-1)^{m_{1}-s})\right]
\end{equation}
The probability vanishes if $m_{1}-s$ is odd. The result for one extra
particle is shown in Fig. \ref{figPairPlusSing}. The results are then very different from those of Fig.
\ref{figNeq51}. If there are two excess particles the steps on the side of the
peak are completely smoothed out. The conclusion of that, in this case, the
model of independent and repeated Hong-Ou-Mandel scatterings does not provide
a good representation of the phenomenon at all.

\section{Conclusion}

A single beam splitter cannot exhibit quantum non-local effects; violating
local realism requires the combination of several such devices to form
interferometers \cite{Laloe-Mullin}.\ Nevertheless, here we have seen one beam
splitter is sufficient to obtain interesting quantum effects, provided it
receives Fock states at its two inputs; these effects are similar to the
\textquotedblleft population oscillations\textquotedblright\ predicted in more
elaborate cases \cite{DBRP, PO}. The oscillations are related to the
Hong-Ou-Mandel effect, but they cannot be understood as a simple juxtaposition
of many separate two-photon-experiments.\ Actually, many-boson effects take
place as a consequence of quantum statistics, which can be understood as a
consequence of the tendency of two Fock states to acquire a relative phase
under the effect of quantum measurement.\ Since this phase is completely
unknown, the characteristic dependence shown in Fig. \ref{figNoLam} results,
onto which quantum oscillations are superimposed. Experimentally, the major
difficulty for observing these effects is the production of Fock states with
well-defined populations.\ Nevertheless, the experimental techniques that have
been developed for Bose-Einstein condensates in ultra-cold gases seem well
suited to planning experiments with input states that contain for instance a
few tens of bosons.

Note added in proofs: the authors have recently become aware of Ref. \cite{CKR}, which gives a study of the localization of phase
obtained by measurements of particles at the output of a single beam splitter. The theoretical treatment is similar to ours but this reference does not assume that all particles are detected; the oscillations we have discussed (generalization of the Hong-Ou-Mandel effect) do not appear.


\begin{thebibliography}{99}                                                                                               %


\bibitem {Rauch-neutron-interferometry}H.\ Rauch, \textquotedblleft Neutron
interferometry\textquotedblright, Science \textbf{262,} 1384 (1993).

\bibitem {Rauch-livre}H.\ Rauch and S.\ Werner, \textquotedblleft Neutron
Interferometry: Lessons in Experimental Quantum Mechanics\textquotedblright,
Clarendon Press (2000).

\bibitem {GHZ}D.M.\ Greenberger, M.A.\ Horne and A.\ Zeilinger,
\textquotedblleft Bell's theorem, quantum theory, and conceptions of the
universe\textquotedblright, M.\ Kafatos ed., Kluwer, p. 69-72, 1989.

\bibitem {GHZ-bis}D.M.\ Greenberger, M.A.\ Horne, A.\ Shimony, A.\ Zeilinger,
``Bell's theorem without inequalities'', Am.\ J.\ Phys. \textbf{58}, 1131-1143 (1990).

\bibitem {GHZ-expts-1}J.W. Pan, D. Bouwmeester, M.\ Daniell, H.\ Weinfurter
and A.\ Zeilinger, \textquotedblleft Experimental test of quantum nonlocality
in three-photon Greenberger--Horne--Zeilinger entanglement\textquotedblright,
Nature \textbf{403}, 515-519 (2000)

\bibitem {Entanglement-swapping}J.W. Pan, D. Bouwmeester, M.\ Daniell,
H.\ Weinfurter and A.\ Zeilinger,\textquotedblleft Experimental Entanglement
Swapping: Entangling Photons That Never Interacted\textquotedblright, Phys.
Rev. Lett. \textbf{80}, 3891--3894 (1998).

\bibitem {Scully}S.\ Prasad, M.O.\ Scully and W.\ Martienssen,
\textquotedblleft A quantum description of the beam splitter\textquotedblright%
, Opt.\ Comm.\ \textbf{63}, 139-145 (1987).

\bibitem {Huttner}B.\ Huttner and Y.\ Ben-Aryeh, \textquotedblleft Influence
of a beam splitter on photon statistics\textquotedblright,
Phys.\ Rev.\ \textbf{A 38}, 204-211 (1988).

\bibitem {Luis}A.\ Luis and L.L. Sanchez-Soto, \textquotedblleft A quantum
description of the beam splitter\textquotedblright, Quantum Semiclass. Opt.
\textbf{7}, 153-160 (1995).

\bibitem {Mullin-Laloe}W.J.\ Mullin and F.\ Lalo\"{e}, \textquotedblleft
Interference of Bose-Einstein condensates: quantum non-local
effects\textquotedblright, Phys.\ Rev.\ \textbf{A 78}, 061605 (2008).

\bibitem {Laloe-Mullin}F.\ Lalo\"{e} and W.J.\ Mullin, \textquotedblleft
Interferometry with independent Bose-Einstein condensates: parity as an
EPR/Bell variable\textquotedblright, Eur.\ Phys.\ J.\ \textbf{70}, 377-396 (2009).

\bibitem {Holland-Burnett}M.J.\ Holland and K.\ Burnett, \textquotedblleft
Interferometric detection of optical phase shifts at the Heisenberg
limit\textquotedblright, Phys.\ Rev.\ Lett.\ \textbf{71}, 1355-58 (1993).

\bibitem {Cosme}O.Cosme, S. Padua, F. Bovino, A. Mazzei, F. Sciarrino and F.
De Martini, \textquotedblleft Hong-Ou-Mandel interferometer with one and two
photon pairs\textquotedblright,; Phys. Rev. \textbf{A 77}, 053822 (2008).

\bibitem {Pritchard}P.J.\ Martin, B.G.\ Oldaker, A.H.\ Miklich and
D.E.\ Pritchard, \textquotedblleft Bragg scattering of atoms from a standing
light wave\textquotedblright, Phys.\ Rev.\ Lett.\ \textbf{60}, 515-518 (1988).

\bibitem {Chu}A.P.\ Chu, K.S.\ Johnson and M.G.\ Prentiss, \textquotedblleft
Atomic beam splitters with achromatic transverse-momentum
transfer\textquotedblright, J.\ Ops.\ Soc.\ Am.\ \textbf{B 13}, 1352-61 (1996).

\bibitem {Sackett}O.\ Garcia, B.\ Deissler, K.J.\ Hughes, J.M.\ Reeves and
C.A.\ Sackett, \textquotedblleft Bose-Einstein-condensate interferometer with
macroscopic separation\textquotedblright, Phys.\ Rev.\ \textbf{A 74},
0311601(R) (2006).

\bibitem {Hong-Ou-Mandel}C. K. Hong, Z. Y. Ou, and L. Mandel, \textquotedblleft
Measurement of subpicosecond time intervals between two photons by interference\textquotedblright,
Phys. Rev. Lett. \textbf{59}, 2044 (1987).

\bibitem {DBRP}J.A.\ Dunningham, K.\ Burnett, R.\ Roth and W.D.\ Phillips, \textquotedblleft
Creation of macroscopic superposition states from arrays of Bose-Einstein condensates\textquotedblright,
New J.\ Phys.\ \textbf{8}, 182 (2006).

\bibitem {PO}W.J.\ Mullin and F.\ Lalo\"{e}, \textquotedblleft Beyond
spontaneously broken symmetry in Bose-Einstein condensates\textquotedblright,
arXiv: 0912.5360.

\bibitem {CKR}H. Cable, P.L. Knight and T. Rudolph, \textquotedblleft
Measurement-induced localization of relative degrees of freedom\textquotedblright,
Phys. Rev. \textbf{A 71}, 042107 (2005).

\end{thebibliography}
\end{document}